\documentclass[aps,twocolumn,floatfix]{revtex4}
\usepackage{float}
\usepackage{graphics,color}
\usepackage{amsmath}
\usepackage{epsfig}
\usepackage{changebar}
\usepackage{footmisc}
\usepackage{epsfig}
\usepackage{amsfonts}
\usepackage{amssymb}
\usepackage{amsbsy}
\usepackage{xcolor}
\usepackage{braket}

\begin{document}

\title{Ordering Kinetics in the Random Bond XY Model}
\author{Manoj Kumar$^1$, Swarnajit Chatterjee$^2$, Raja Paul$^2$\footnote{author for correspondence: ssprp@iacs.res.in} and Sanjay Puri$^1$\footnote{author for correspondence: purijnu@gmail.com}}

\affiliation{$^1$ School of Physical Sciences, Jawaharlal Nehru University, New Delhi -- 110067, India.}
\affiliation{$^2$ Indian Association for the Cultivation of Science, Kolkata -- 700032, India.}

\begin{abstract}
We present a comprehensive Monte Carlo study of domain growth in the random-bond XY model with non-conserved kinetics. The presence of quenched disorder slows down domain growth in $d=2,3$. In $d=2$, we observe power-law growth with a disorder-dependent exponent on the time-scales of our simulation. In  $d=3$, we see the signature of an asymptotically logarithmic growth regime. The scaling functions for the real-space correlation function are seen to be independent of the disorder. However, the same does not apply for the two-time autocorrelation function, demonstrating the breakdown of superuniversality.    
\end{abstract}

\maketitle
\section{Introduction}

The XY model has been widely studied in the literature. Experimentally, a large number of physical systems have been described by the XY model. For dimensionality $d=2$, typical realizations of the XY model include magnetic films with planar anisotropy~\cite{BR}, thin-film superfluids or superconductors~\cite{egk}, Josephson junction arrays~\cite{hf,bmo}, hexatic liquid crystals~\cite{pgy}, melting of two-dimensional solids~\cite{kjs}, etc.  In $d=3$, physical systems such as superfluid $^4$He~\cite{hm,gh} and planar spin magnets have been described by the XY model.  

The XY model in  $d=2$  exhibits the  well-known {\it Berezinskii-Kosterlitz-Thouless} (BKT) transition at temperature $T_{\rm BKT}$~\cite{beren,kt,koster}. This system shows long-range order (LRO) only at temperature $T=0$. However, for $0<T<T_{\rm BKT}$, the system shows quasi-long-range order (QLRO), where the correlation-function decays as a power-law with temperature-dependent exponent $\eta(T)$. In this state, the morphology consists of bound states of vortex-antivortex pairs. For $T>T_{\rm BKT}$, the vortex-antivortex pairs unbind and the correlation function decays exponentially. In this paper, we are interested in the disordered XY model. The presence of quenched disorder has a strong effect on the BKT phase transition. For instance, various numerical studies~\cite{ty,lcp,alono,mb,harris,wpm} of the XY model with dilution (e.g., site-vacancies, bond-vacancies) have shown that $T_{\rm BKT}$ decreases with increasing disorder. It becomes zero at a critical value of dilution, which is referred to as the  percolation threshold. In $d=3$, the disorder-free XY  model exhibits true long-range order (LRO) for $T<T_c$~\cite{ksw}.

In this paper, we are interested in the nonequilibrium ordering kinetics of the disordered XY model in $d=2,3$, subsequent to a quench from high $T$. When a pure XY system is quenched to $T<T_{\rm BKT}$ or $T<T_c$, the coarsening process is characterized by the annihilation of vortex-antivortex pairs~\cite{pr, puri92}. The characteristic length scale $R(t)\sim t^{1/2}$ for non-conserved vector fields in $d\geq3$~\cite{bp}. For $d=2$, Yurke et al.~\cite{ypkh} predicted a  logarithmic correction to the diffusive growth as $R(t)\sim (t/ \ln t)^{1/2}$. 

In recent years, there has been intense interest in the subject of domain growth in disordered systems~\cite{pcp,pp,ppr}. In general, the domain boundaries become trapped at late times by  energy barriers introduced by the disorder, thereby slowing down the asymptotic domain growth law~\cite{puri04}.
There has been some debate about the precise nature of the asymptotic growth law in the case with scalar order parameter, e.g., {\it random-field Ising model} (RFIM) or {\it random-bond Ising model} (RBIM). The early Monte Carlo (MC) simulations of the RBIM by Paul et al.~\cite{ppr} reported a power law growth with a disorder-dependent exponent. However, the recent works of Corberi et al.~\cite{lmpz,clmpz2011,clmpz} have demonstrated that there is a slow crossover to a logarithmic growth regime, which is numerically very difficult to access. 

It would be fair to say that we now have a good understanding of the case with scalar order parameter. However, to the best of our knowledge, there have been no studies of the case with vector order parameter -- even though this is experimentally very important. This paper is a first step in that direction, i.e., we present comprehensive numerical results for domain growth in the {\it random-bond XY model} (RBXYM) with nonconserved kinetics in $d=2,3$. Our study covers two important aspects. First, we study the effect of disorder on the transition temperature by using the Wolff single-cluster updating algorithm~\cite{ulli}. Second, we study nonconserved ordering kinetics via the Metropolis algorithm~\cite{mrrt} by quenching the system below the transition temperature. The main results of our study are as follows: \\
(a) The critical temperature is found to decrease with increasing disorder. \\
(b) For domain growth, the correlation function shows dynamical scaling in $d=2, 3$. Also, the scaling function is independent of disorder, and therefore shows a universal behavior. However, the two-time autocorrelation function is not universal. \\
(c) In $d=2$, the growth law over our simulation time-scales  is algebraic with a disorder-dependent exponent. \\ 
(d) In $d=3$, the domain growth law is asymptotically logarithmic, as in the scalar case.

This paper is organized as follows. In Sec.~\ref{s2}, we discuss the model and present  details of our numerical simulations. In Sec.~\ref{s3}, we present detailed numerical results from our simulations of the $d=2,3$ RBXYM. Finally, in Sec.~\ref{s4}, we conclude this paper with a summary and discussion of the results.

\section{Modeling and Simulation Details}
\label{s2}

\subsection {Random Bond XY model}

Consider a lattice of size $L^2$ (in $d=2$) or $L^3$ (in $d=3$). Each lattice site labeled by $i$ has a two-component vector spin $\boldsymbol{S}_i$. The Hamiltonian for the RBXYM is defined as 
\begin{eqnarray}
\label{refH}
\mathcal{H} &=& -\sum_{\langle ij \rangle} J_{ij} \boldsymbol{S}_i \cdot \boldsymbol{S}_j \nonumber \\
&=& -\sum_{\langle ij \rangle} J_{ij} \cos(\theta_i-\theta_j),
\end{eqnarray}

where $J_{ij}$ is the exchange coupling between the nearest-neighbor pair denoted by ${\langle ij \rangle}$. Each spin $\boldsymbol{S}_i=(\cos \theta_i,\sin \theta_i)$ is a unit vector and is described by an angle $\theta_i \in (-\pi,\pi)$. The quenched random-bond variables $\{J_{ij}\}$ are distributed uniformly on the interval $[1-\epsilon/2,1+\epsilon/2]$, where $\epsilon$ quantifies the degree of disorder. The limit $\epsilon=0$ corresponds to the pure case ($J_{ij}=1$). Here,  we focus on the ferromagnetic case where $J_{ij}>0$. Therefore, $\epsilon=2$ corresponds to the maximum value of disorder in our simulations.

\subsection{Simulation Details for Study of Transition Temperature}
\label{sim_tc}

Before we discuss nonequilibrium studies, it is important to understand the equilibrium properties of the RBXYM. In this context, let us discuss the simulation details for determining the transition temperature $T_{\rm BKT}$ (in $d=2$) or $T_c$ (in $d=3$) in the presence of disorder ($\epsilon$). For the pure XY model, the transition temperature is $T_{\rm BKT} \simeq 0.89$ in $d=2$ \cite{tc,ffs}, and $T_c \simeq 2.203$ in $d=3$ \cite{hm,fmw}.

A standard tool to determine the transition temperature is the fourth-order Binder cumulant $U_4 (T, L)$ \cite{bin,loison}, defined  as
\begin{equation}
\label{u4}
U_4 (T, L)= 1-\frac{[\langle m^4 \rangle]}{3[\langle m^2 \rangle ^2]}.                               
\end{equation}
Here, $m$ is the magnetization, and $\langle \cdots \rangle$ and $[\cdots]$ denote the thermal and disorder averages, respectively. The Binder cumulants $U_4 (T, L)$ are plotted against temperature $T$ for different lattice sizes $L$. Then, in the scaling region near $T_{\rm BKT}$ or $T_c$, the $U_4$ vs. $T$ curves for different $L$ have a unique intersection point~\cite{bin}, which is identified as the transition temperature.

The magnetization $m$ for the XY model is defined as 
\begin{equation}
m=\frac{1}{N} \sqrt{\left(\sum_{i=1}^N\cos\theta_i\right)^2+\left(\sum_{i=1}^N\sin\theta_i\right)^2}  ,
\end{equation}
where $N$ is the total number of sites, i.e., $N=L^d$. The magnetization $m$ is measured when the system has reached  thermal equilibrium. To equilibrate the system at temperature $T$, we use the canonical sampling MC method with the Wolff single-cluster updating algorithm \cite{ulli}.

A single MC update in the Wolff single-cluster algorithm can be described as follows: \\
(a) Choose a random reflection   $\boldsymbol{r} =(\cos {\phi},\sin {\phi})$ and a random spin $\boldsymbol{S}_i=(\cos \theta_i,\sin \theta_i)$ as the starting point for a cluster $\mathcal{C}$ to be built. \\ (b) Flip $\boldsymbol{S}_i\rightarrow R(\boldsymbol{r})\boldsymbol{S}_i =  \boldsymbol{S}_i-2(\boldsymbol{S}_i\cdot\boldsymbol{r})\boldsymbol{r}$, i.e., $\theta_i\rightarrow \theta_i'=\pi-\theta_i+2\phi$. \\
(c) Visit all neighboring spins $\boldsymbol{S}_j$ of $\boldsymbol{S}_i$, and add them to the cluster  $\mathcal{C}$ with the probability \cite{ulli} 
\begin{equation}
\label{refP}
P(\boldsymbol{S}_i, \boldsymbol{S}_j)= 1-\exp \{\text{min} [0, 2 \beta J_{ij}(\boldsymbol{r} \cdot \boldsymbol{S}_i)(\boldsymbol{r} \cdot \boldsymbol{S}_j)] \} ,
\end{equation}
where $\beta=(k_B T)^{-1}$. In terms of the angle variables, the corresponding expression for the probability is
\begin{equation}
P(\theta_i,\theta_j)= 1-\exp \{\text{min} [0, 2 \beta J_{ij} \cos(\theta_i-\phi)\cos(\theta_j-\phi)] \} .
\end{equation}
(d) Keep visiting all nearest neighbors of newly-added spins, and add them to the cluster  $\mathcal{C}$ with probability $P$. Continue this process until no spin is left to add to $\mathcal{C}$.  One Monte Carlo step (MCS) corresponds to $N$ such updates.

\subsection{Simulation Details for Study of Ordering Kinetics}
\label{sim_oks}

We study ordering kinetics in the RBXYM by assigning a random initial orientation to each spin $\theta_i \in (-\pi,\pi)$, mimicking the high-temperature disordered state. At time $t=0$, the system is rapidly quenched to $T < T_{\rm BKT}$ (in $d=2$) or $T < T_c$ (in $d=3$), and evolved via nonconserved kinetics. We let the system evolve upto $10^6$ MCS with the help of the Metropolis algorithm \cite{mrrt}. For the  XY system, our algorithm can be described as follows: \\
(a) Select a random spin $\boldsymbol{S}_i$ and give $\theta_i$ a small rotation $\delta \in (-0.1,0.1)$, i.e., $\theta_i\rightarrow\theta_i'= \theta_i + \delta$. \\
(b) The new spin $\theta_i'$ is accepted with the Metropolis transition probability 
\begin{equation}
P=\text{min}[1,\exp\,(-\beta\varDelta\mathcal{H})] . 
\end{equation}
Here, $\varDelta\mathcal{H}$ is the change in energy resulting from the angle change $\theta_i\rightarrow \theta_i'$:
\begin{equation}
\Delta\mathcal{H}=\sum_{k}J_{ik} \left[\cos(\theta_i-\theta_{k})-\cos(\theta_i'-\theta_{k})\right] ,
\end{equation}
where $k$ refers to the nearest neighbors of site $i$.

A useful quantity in studying phase ordering kinetics is the {\it spatial correlation function}, which is defined as~\cite{puri} 
\begin{equation}
\label{refC}
C({\textbf{r}},t)=\frac{1}{N}\sum_{i=1}^{N} \left\{ \left[ \langle \boldsymbol{S}_i(t) \cdot \boldsymbol{S}_{i+\textbf{r}}(t) \rangle \right] - \left[\langle \boldsymbol{S}_i(t) \rangle \right] \cdot \left[\langle \boldsymbol{S}_{i+\textbf{r}}(t) \rangle \right] \right\} ,
\end{equation}
where $\left[\langle \cdots \rangle \right]$ indicates an averaging over independent runs, and different realizations of bond randomness. The quantity  $C({\textbf{r}},t)$ characterizes the morphology of the coarsening system. If the system is isotropic and characterized by a single length scale $R(t)$, then the correlation function has a dynamical scaling form \cite{puri,bray94}$\colon$
\begin{equation}
\label{refScaling1}
C({\textbf{r}},t) = f\left(\frac{r}{R(t)}\right),
\end{equation} 
where $f(x)$ is the scaling function. The characteristic length scale $R(t)$ is defined as the distance over which  the correlation function decays to (say) 0.2 of its maximum value. In the XY model, the typical length-scale can also be determined from the density of topological defects, e.g., $\rho_{\rm def}(t)\sim1/R_v(t)^2$ in $d=2$, where $R_v$ is the diameter of a vortex. In the scaling regime, this definition will differ from the former one only by a prefactor. In this paper, we use the $R(t)$ determined from the decay of $C(r,t)$~\cite{rr,bb}.

Another nonequilibrium quantity of interest is the {\it two-time autocorrelation function}:
\begin{equation}
\label{autoC}
A(t,t_w)=\frac{1}{N}\sum_{i=1}^{N} \left\{ \left[\langle \boldsymbol{S}_i(t_w) \cdot \boldsymbol{S}_i(t) \rangle \right] - \left[ \langle \boldsymbol{S}_i(t_w) \rangle \right] \cdot \left[ \langle \boldsymbol{S}_{i}(t) \rangle \right] \right\} ,
\end{equation}
which is important in studies of aging \cite{mz09}. The quantity $t_w$ is referred to as the {\it waiting time}. In the scaling regime, we expect
\begin{equation}
 A(t,t_w)=h\left[\frac{R(t)}{R(t_w)}\right]\sim \left[\frac{R(t)}{R(t_w)}\right]^{-\lambda} \quad \text{for}\,\, t\gg t_w .
\end{equation}
Here, $h(y)$ is a scaling function, and the exponent $\lambda$ was first introduced in the context of spin glasses~\cite{fh88}.

The morphology of the ordering system is usually studied by scattering experiments which measure the structure factor $S(\textbf{k},t)$, defined as the Fourier transform of the correlation function $C({\textbf{r}},t)$. It has the corresponding dynamical scaling form:
\begin{equation}
S(k, t)=R(t)^{d}g(kR(t)) . 
\end{equation}
Bray and Puri (BP) \cite{bp} and Toyoki (T) \cite{toyoki} have independently obtained the scaling functions $f(x)$ and $g(p)$ for domain growth with an $n$-component vector field without disorder. They found that the scaling function $g(p)$ has the large-$p$ behavior:
\begin{equation}
g(p) \sim p^{-(d+n)}~~~\mbox{for}~~~ p \rightarrow \infty .
\end{equation}
This is referred to as the {\it generalized Porod tail}, as Porod~\cite{porod,op88} emphasized that scattering off sharp interfaces in a scalar field ($n=1$ case) yields the structure-factor tail $g(p) \sim p^{-(d+1)}$.

\section{Numerical Results}
\label{s3}

Now, we will present  numerical results from our simulations of the $d=2,3$ RBXYM. We consider the disorder values $\epsilon=0$, 0.5, 1.0, 1.5, 2.0; $\epsilon=0$ is the pure case for reference, and $\epsilon=2$ corresponds to the case with maximum disorder. First, we determine the transition temperatures as a function of $\epsilon$. Then, we study ordering kinetics by quenching the system below $T_{\rm BKT} (\epsilon)$ or $T_c(\epsilon)$.

\subsection{RBXYM in $d=2$}

\subsubsection{Estimation of $T_{\rm BKT}$}

Let us first consider the $d=2$ RBXYM. We study this system on a square lattice ($L^2$) of linear sizes $L$ = $96$, $128$, and $256$.  Starting from a random initial configuration, we let the system equilibrate using the Wolff single-cluster update algorithm (see Sec.~\ref{sim_tc}). After equilibration, we average $m^2$ and $m^4$ upto $6 \times 10^5$ MCS. Further, we perform a disorder average over 200 independent runs of random-bond configurations. Then, one can determine the Binder cumulant $U_4(T, L)$ from Eq.~\eqref{u4}.

\begin{figure}[htbp]
\includegraphics[width=\columnwidth]{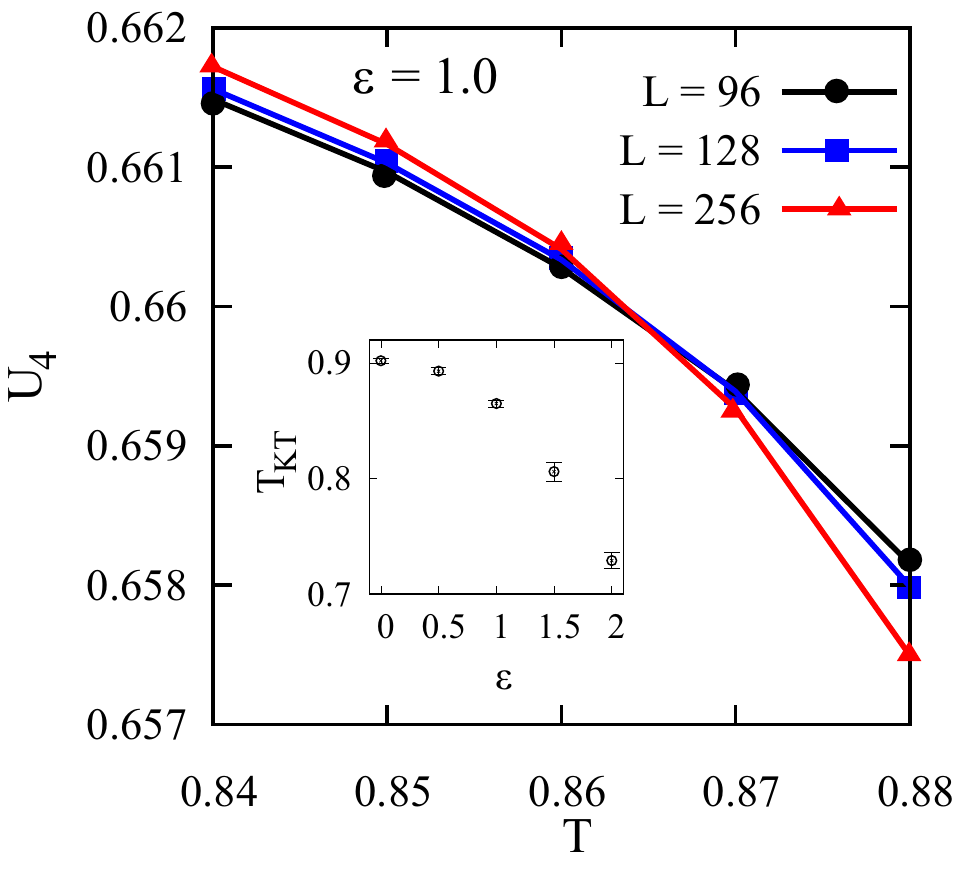}\\
\caption{Plot of fourth-order Binder  cumulant $U_4(T,L)$ vs. temperature $T$ for the $d=2$ RBXYM. The data is shown for $\epsilon$ = 1.0, and square lattices with $L =96, 128,256$. The transition temperature $T_{\rm BKT}(\epsilon)$ is determined from the intersection of different $L$-curves. The inset shows the plot of $T_{\rm BKT}(\epsilon)$ with disorder $\epsilon$. The numerical values of $T_{\rm BKT}(\epsilon)$ are provided in Table~\ref{table1}.}
\label{fig1}
\end{figure}

\begin{table}[htbp]
\begin{center}
\begin{tabular}{|p{0.05\textwidth}|c|c|}
\hline
$\epsilon$ & $T_{\rm BKT}$ & $\overline{z}$ \\
\hline
0.0 & 0.902 $\pm$ 0.002 & 2 \\
\hline
0.5 & 0.893 $\pm$ 0.003 & 2.09 $\pm$ 0.01 \\
\hline
1.0 & 0.865 $\pm$ 0.003 &  2.16 $\pm$ 0.02 \\
\hline
1.5 & 0.806 $\pm$ 0.008 &  2.29 $\pm$ 0.04 \\
\hline
2.0 & 0.729 $\pm$ 0.007 &  2.43 $\pm$ 0.03 \\
\hline
\end{tabular}
\caption{Transition temperatures $T_{\rm BKT}(\epsilon)$ and growth exponents $\overline{z}(\epsilon)$ for the $d=2$ RBXYM.} 
\label{table1}
\end{center}
\end{table}

Figure~\ref{fig1} is a plot of $U_4$ vs. $T$ for $\epsilon$ = 1.0. (We have zoomed the plot in the vicinity of $T_{\rm BKT}$.) The transition temperature ($T_{\rm BKT}$) can be accurately  estimated from the intersection of Binder cumulant curves for different $L$. The $T_{\rm BKT}$-values for various $\epsilon$ are plotted in the inset of Fig.~\ref{fig1}, and tabulated in Table~\ref{table1}. For the pure case ($\epsilon=0$), we found $T_{\rm BKT}=0.902\pm 0.002$, which is consistent with the value $T_{\rm BKT}\simeq 0.893$ in the literature ~\cite{kt,koster,tc,ffs}. With increasing $\epsilon$, $T_{\rm BKT}$ decreases from $T_{\rm BKT}\simeq 0.902$ to $T_{\rm BKT}(\epsilon=2)= 0.729 \pm 0.007$. 

\subsubsection{\emph{Coarsening dynamics}}

Next, we present numerical results for  coarsening dynamics in the $d=2$ RBXYM. The simulations are performed on a square lattice of size  $1024^2$ with periodic boundary conditions applied on both sides. As described in Sec.~\ref{sim_oks}, the initially disordered system is quenched to $T=0.2$ or $0.5$ ($<T_{\rm BKT}$, see Table~\ref{table1}) at $t=0$ MCS. The system is evolved upto $t$ = $10^{6}$ MCS using the Metropolis algorithm.  All statistical results presented here are averaged over 20 runs (sometimes more) with independent $\{J_{ij}\}$-configurations.

\begin{figure}[htbp]
\includegraphics[width=\columnwidth]{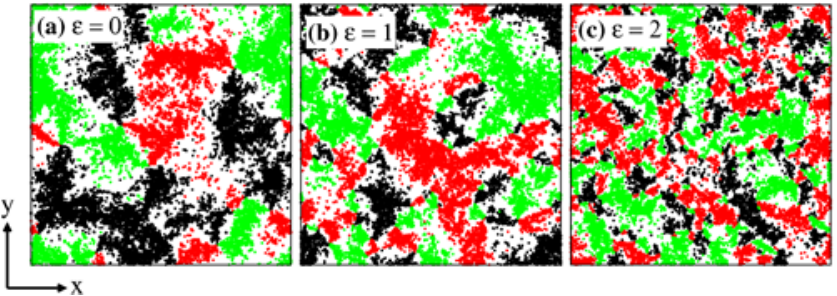}\\
\caption{Evolution snapshots of the $d=2$ RBXYM at $t = 10^{6}$ MCS, after a quench from $T$ = $\infty$ to $T$ = $0.2$  for (a) $\epsilon= 0$, (b) $\epsilon= 1$,  and (c) $\epsilon= 2$. The lattice size is $512^2$. In these  plots, $\{\theta_i\}$ are marked in the interval $[\theta_0 - 0.1, \theta_0 + 0.1]$ with the following color coding: $\theta_0 =0$ (black), $\theta_0=2\pi/3$ (red/gray), $\theta_0 = - 2 \pi/3$ (green/light gray).}
\label{fig2}
\end{figure}

In Fig.~\ref{fig2}, we show the typical evolution snapshots for $T=0.2$. The snapshots correspond to $t$ = $10^{6}$ MCS,  and  disorder  amplitudes $\epsilon=0, 1, 2$. The three colors denote small angle-windows, as specified in the caption. A junction point of the three colors corresponds to a vortex or anti-vortex, depending on the direction of rotation.  These plots  show an increase  in defect density  with disorder, corresponding to slowing down of domain growth.

\begin{figure}[htbp]
\includegraphics[width=\columnwidth]{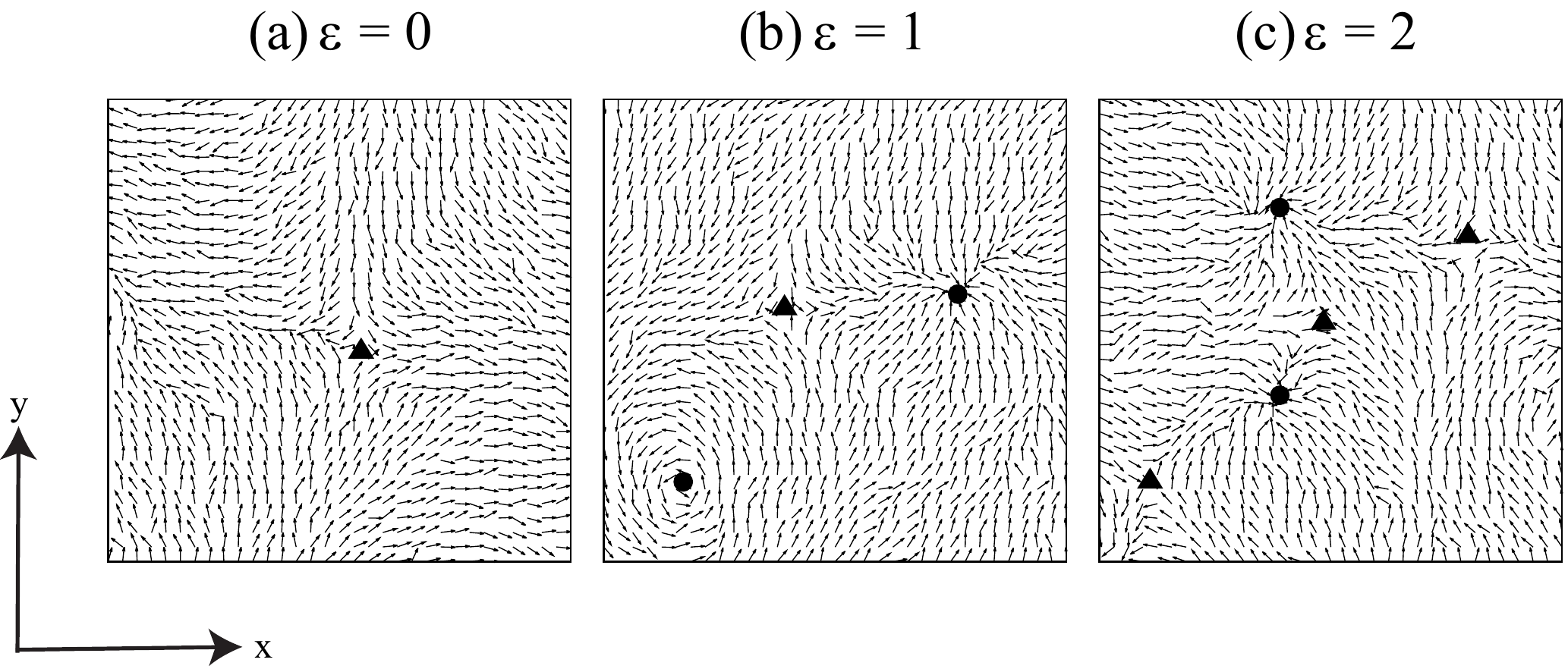}\\
\caption{Vector plots for $\{\theta_i\}$-configurations in Fig.~\ref{fig2}. At each lattice site $i$, we draw a vector corresponding to $\boldsymbol S_i=(\cos \theta_i, \sin \theta_i)$. For a better view, we show only a $32^2$ corner of the $512^2$ lattice. Solid circles denote vortices, and solid triangles denote antivortices.}
\label{fig3}
\end{figure}

In Fig.~\ref{fig3}, we show vector plots corresponding to Fig.~\ref{fig2}, in which we draw a unit vector for each spin $\boldsymbol{S}_i=(\cos \theta_i,\sin \theta_i)$. For a better visualization of vectors, we have shown only a  small portion of the lattice in Fig.~\ref{fig2}. Vortices and anti-vortices are marked by solid circles and triangles, respectively. These are characterized by calculating the net change in spin direction while moving clockwise on a square plaquette. A vortex is identified if a spin rotates through $2\pi$, and  an anti-vortex is identified if a spin rotates through $-2\pi$.

\begin{figure}[htbp]
\includegraphics[width=\columnwidth]{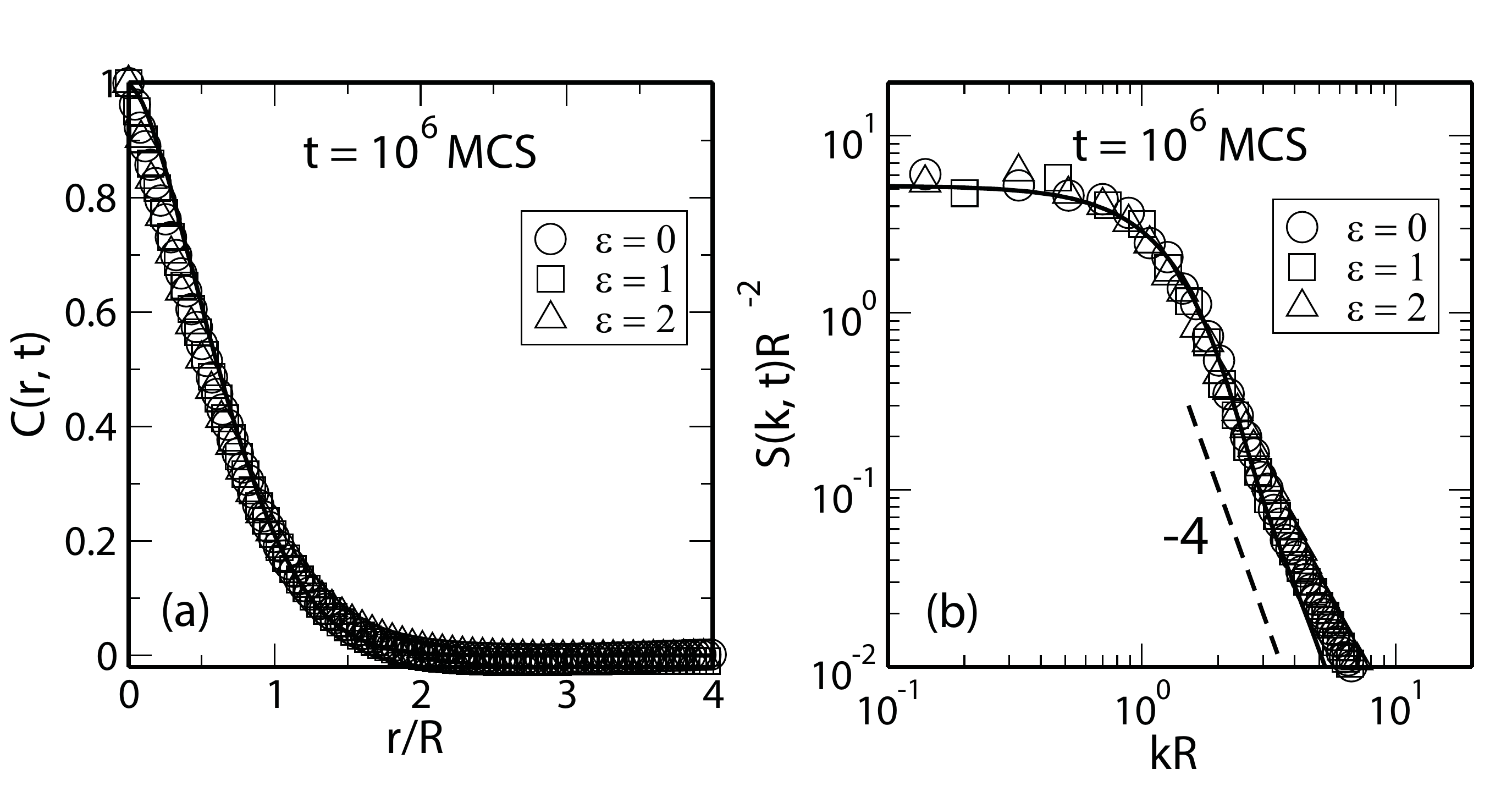}\\
\caption{(a) Scaled correlation functions, $C(r,t)$ vs. $r/R(t)$, for the evolution of the $d=2$ RBXYM after a quench to $T=0.5$. We show data for $t = 10^{6}$ MCS, and $\epsilon=0,1,2$. (b) Scaled structure factors, $S(k,t)R(t)^{-2}$ vs. $kR$, for the data sets in (a). The solid curves in (a) and (b) denote the Bray-Puri-Toyoki (BPT) function in Eq.~\eqref{bpt} for $n = 2$, and its Fourier transform, respectively. The line of slope $-4$ in (b) denotes the generalized Porod law: $S(k,t) \sim k^{-(d+n)}$ for $d = n = 2$.}
\label{fig4}
\end{figure}

In Fig.~\ref{fig4}, we plot the scaled forms of (a) the correlation function, $C(r,t)$ vs. $r/R$; and (b) the structure factor, $S(k,t)R^{-2}$ vs. $kR$. The data sets correspond to different values of $\epsilon$. We have confirmed (not shown here) that the data sets for a fixed value of $\epsilon$ and different times show a good data collapse. Thus, dynamical scaling holds for each value of $\epsilon$. In Fig.~\ref{fig4}, we check for the robustness of this scaling function by plotting scaled data at $t$ = $10^{6}$ MCS for $\epsilon=0,1,2$. The scaling functions are seen to be independent of the disorder amplitude. This feature was first seen in simulations of ordering kinetics in the RBIM~\cite{pcp,pp}, and was referred to as super-universality (SU). We will shortly see that the SU property does not apply to the two-time function $A(t,t_w)$. 

The solid curve in Fig.~\ref{fig4}(a) is a plot of the Bray-Puri-Toyoki (BPT) function~\cite{bp,toyoki} for $n=2$. As mentioned earlier, BPT obtained the scaling function $f(x)$ for ordering dynamics of the $\mathcal{O}(n)$ model:
\begin{equation}\label{bpt}
f_{\rm BPT}(r/R)=\frac{n\gamma}{2\pi}\left[B\left(\frac{n+1}{2},\frac{1}{2}\right)\right]^2 F\left(\frac{1}{2},\frac{1}{2};\frac{n+2}{2};\gamma^2\right),
\end{equation}
where $\gamma=\exp(-r^2/R^2)$. In Eq.~\eqref{bpt}, $B(x,y)\equiv \Gamma(x)\Gamma(y)/\Gamma(x+y)$ is the beta function, and $F(a,b;c;z)$ is the hypergeometric function. In Fig.~\ref{fig4}(b),  the solid curve is the Fourier transform of the BPT function, and the line of slope $-4$ denotes the generalized Porod law: $S(k,t)\sim k^{-(d+n)}$ for $d=2, n=2$. 

\begin{figure}[htbp]
\includegraphics[width=\columnwidth]{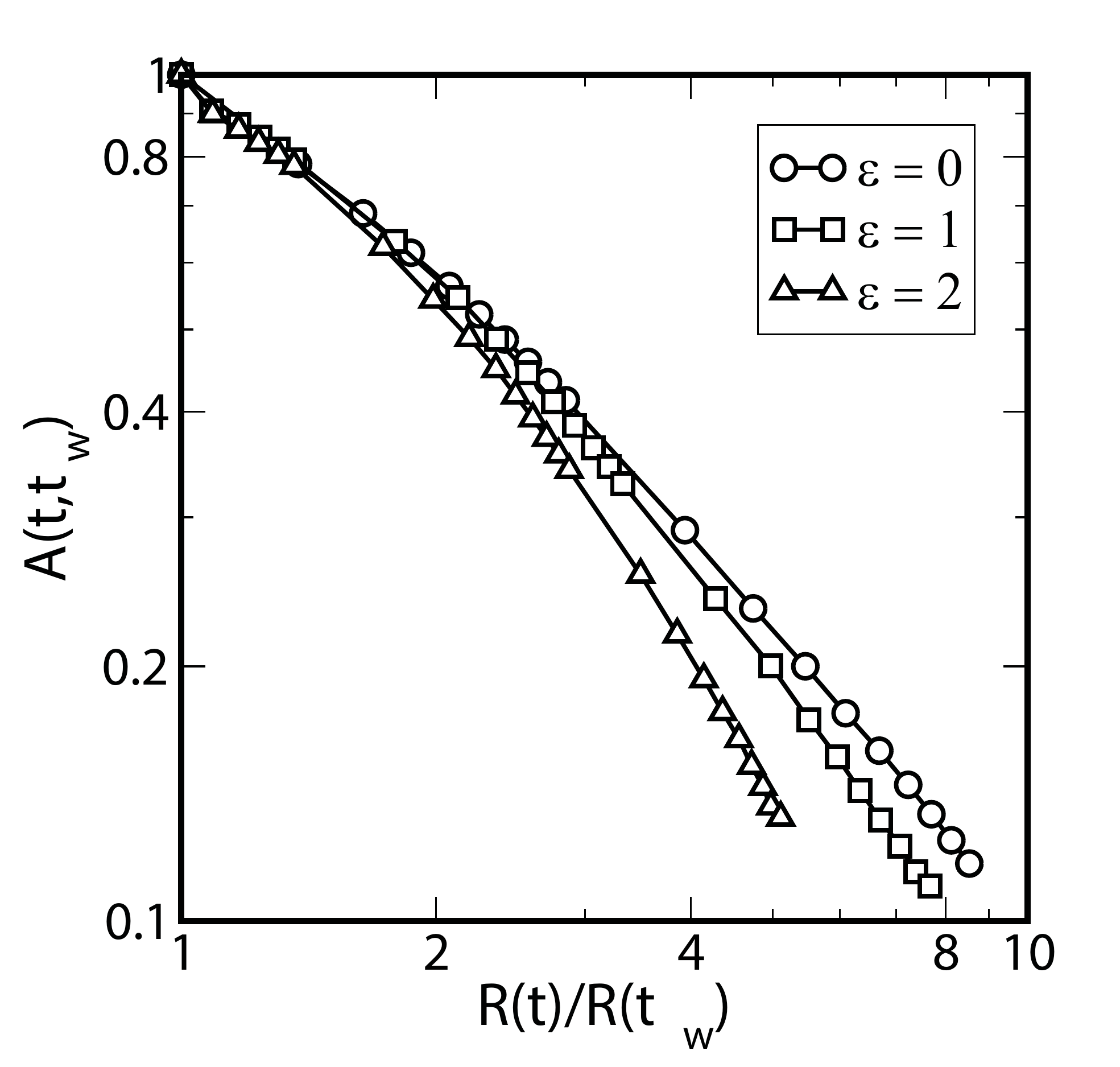}\\
\caption{Plot of the autocorrelation function, $A(t,t_w)$ vs. $R(t)/R(t_w)$, for waiting time $t_w=10^4$ MCS, and $\epsilon = 0,1,2$.}
\label{fig5}
\end{figure}  
 
In Fig.~\ref{fig5}, we examine the scaling properties of $A(t,t_w)$. First, for a fixed value of $\epsilon$, we superpose data for $A(t,t_w)$ vs. $R(t)/R(t_w)$ from different times (not shown here). These data sets show a good scaling collapse.  In Fig.~\ref{fig5}, we plot the corresponding data sets for $t_w=10^4$ MCS and $\epsilon=0,1,2$ -- analogous to our plots in Fig.~\ref{fig4}. In this case, the scaling functions show a clear dependence on the disorder amplitude, demonstrating that SU does not apply for $A(t,t_w)$. The decay exponent $\lambda$ is determined from the asymptotically linear portion of the log-log plot in Fig.~\ref{fig5}. Clearly, $\lambda$ is larger for higher values of $\epsilon$. A similar observation was made earlier in the context of the RBIM with nonconserved kinetics \cite{lmpz,clmpz2011}.

The most important characteristic of a coarsening process is the growth law of $R(t)$. In Figs.~\ref{fig2} and~\ref{fig3}, we have seen that the evolution occurs via the annihilation of vortices and anti-vortices with typical size $R(t)$. It is useful to review the arguments for the growth law in the pure XY model before examining data for the disordered case. 
  
Consider a single vortex-antivortex pair separated by a distance $R$, with $a$ as the dimension of the vortex-core. For a $n$-component vector model in $d$ dimensions,  the topological defects lie on a surface of dimension $d-n$. Therefore, the volume of the defect-core scales as $R^{d-n}$~\cite{bray94,rb}.  For the XY model ($n=2$) in $d=2$, the core volume is a dimensionless constant. The defect pair energy $E_p(R) \sim \ln(R/a)$~\cite{bray94,rb,bray}. Thus, the driving force [$F(R) = -dE_p/dR$]  per unit core-volume, which is  responsible for the annihilation of the vortex-antivortex pair, scales as $F(R)\sim-1/R$.

The annihilation time of the defect pair is governed by the vortex mobility $\mu$, which depends logarithmically on the pair separation, i.e., $\mu \sim [\ln(R/a)]^{-1}$~\cite{bray94,rb,bray}. As the mobility is related to the velocity via $v =\mu F$, we have 
\begin{equation}
\frac{dR}{dt}\sim - \frac{1}{R\ln(R/a)}.
\end{equation}
For a pair separated by a large distance $R \gg a$, integrating this equation gives the annihilation time $t\sim R^2\ln(R/a)$. This  can be inverted to obtain 
\begin{equation}\label{GL1}
 R \sim \left[\frac{t}{\ln(t/a^2)}\right]^{1/2}.
\end{equation}
Eq.~\eqref{GL1} yields the growth law $R(t)\sim (t/ \ln t)^{1/2}$ for the pure  XY model in $d=2$. 
 
\begin{figure}[htbp]
\includegraphics[width=\columnwidth]{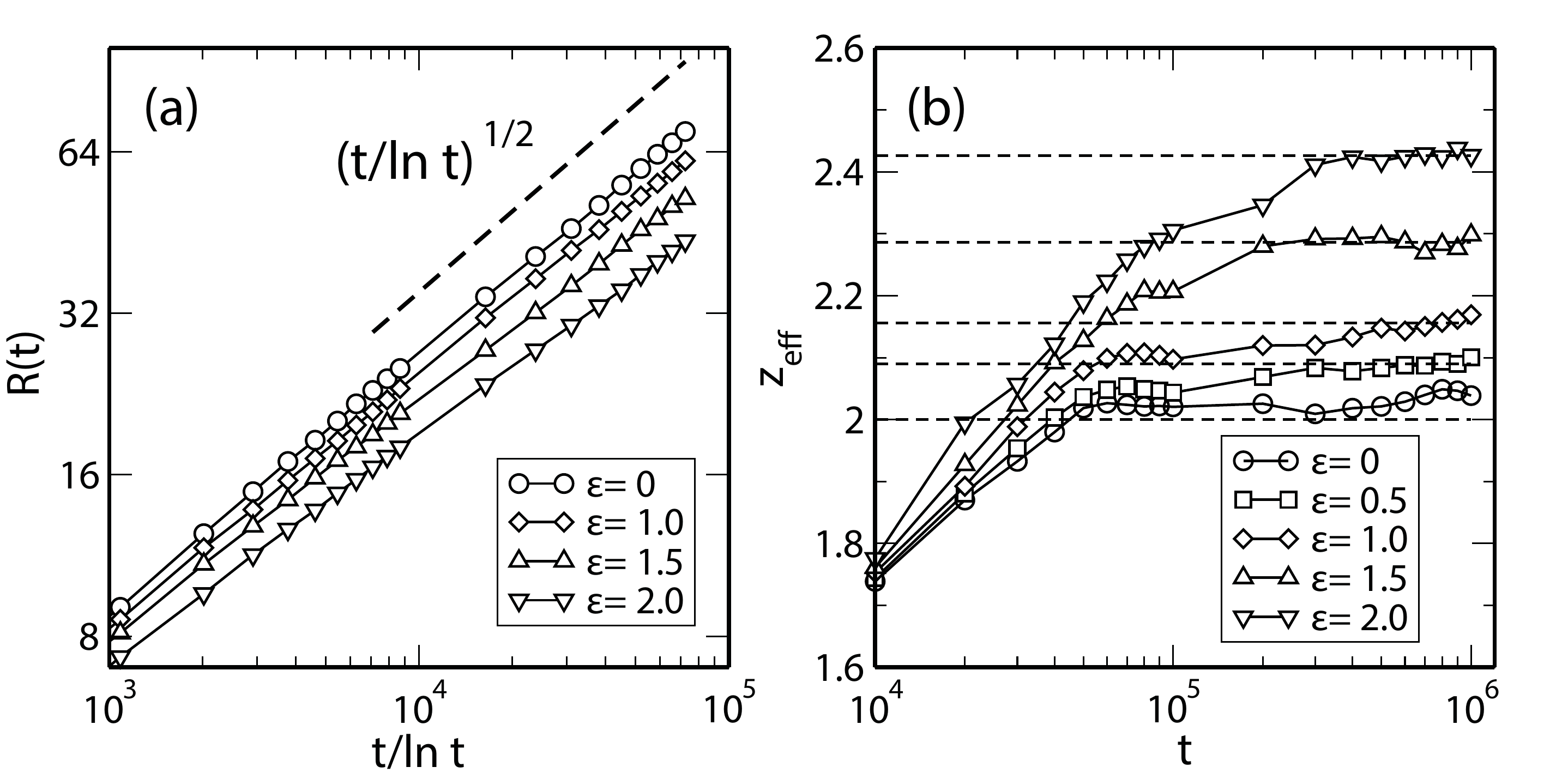}\\
\caption{(a) Plot of $R(t)$ vs. $t/\ln t$ on a log-log scale for the specified values of $\epsilon$.  The dashed line is of slope 0.5, which indicates the growth law for the pure case: $R(t)\sim (t/\ln t)^{1/2}$. (b) Plot of the effective exponent, $z_{\rm eff}$ = $[d(\ln R)/d(\ln [t/\ln t])]^{-1}$ vs. $t$, for $\epsilon = 0,0.5,1.0,1.5,2.0$. The dashed lines denote $\overline{z}(\epsilon)$.}
\label{fig6}
\end{figure}

What is the effect of random-bond disorder on this growth law? As a reference point, it is useful to recall the scenario for the RBIM \cite{ppr,lmpz,clmpz2011}. In that case, coarsening interfaces are trapped by disorder sites with energy barriers which have a power-law dependence on the length scale $R$: $E_B(R) \sim R^\varphi$, where $\varphi$ is the barrier exponent. This yields an asymptotically logarithmic growth regime: $R(t) \sim (\ln t)^{1/\varphi}$, which is preceded by a power-law regime where the exponent depends on the disorder amplitude.  For the $d=2$ RBXYM, Fig.~\ref{fig6}(a) shows the plot of $R(t)$ vs. $t/\ln t$ on a log-log scale for different  $\epsilon$-values. This plot is motivated by the logarithmic correction in the domain growth law for the pure case.  The dashed line denotes the power-law growth for the pure-case: $R(t) \sim (t/\ln t)^{1/2}$ \cite{rr,rb,br}. Figure~\ref{fig6}(a) shows that the presence of disorder slows down domain growth. The data sets in Fig.~\ref{fig6}(a) suggest a power-law (over three decades of $t$) with a disorder-dependent exponent:
\begin{equation}
R(t)\sim \left(\frac{t}{\ln t}\right)^{\phi (\epsilon)} \simeq \left(\frac{t}{\ln t}\right)^{1/\bar{z} (\epsilon)} .
\end{equation}
Before proceeding, we should stress that the log-log plot of $R(t)$ vs. $t$ on the same time-window is also consistent with a power-law behavior. The only difference from Fig.~\ref{fig6}(a) is that the effective exponent $\phi (\epsilon)$ is reduced due to the logarithmic correction. We need at least five decades of data to differentiate between $t^{\bar{\phi}}$ and $(t/\ln t)^\phi$ on a log-log scale.

For a quantitative study of the growth law, we determine the effective growth exponent, defined as
\begin{equation}\label{zeff_xy}
\frac{1}{ z_{\rm {eff}}} = \frac{d\,[\ln R(t)]}{d\left[\ln \left(t/\ln t\right)\right]}.
\end{equation}
In Fig.~\ref{fig6}(b), we plot $z_{\rm {eff}}$ vs. $t$ for the data in Fig.~\ref{fig6}(a). This plot clearly shows an extended flat regime upto the time-scale ($10^6$ MCS) of our simulation. The dashed lines denote the corresponding values of the effective exponent $\overline{z}(\epsilon)$, which are specified in Table~\ref{table1}. This power-law behavior is consistent with the RBIM results at intermediate times~\cite{ppr,lmpz,clmpz2011}. In the RBIM studies of Lippiello et al.~\cite{lmpz,clmpz2011}, there is an upward curvature of the $z_{\rm eff}$ vs. $R$ plot at late times, signaling the onset of the logarithmic regime. We do not clearly see this signature in Fig.~\ref{fig6}(b) for the $d=2$ RBXYM. Clearly, even longer simulations are needed to determine whether the RBXYM shows a crossover to an asymptotic logarithmic regime. This may be conjectured, as we expect the trapping energy of a vortex in the RBXYM to scale with the vortex size.

\begin{figure}[htbp]
\includegraphics[width=\columnwidth]{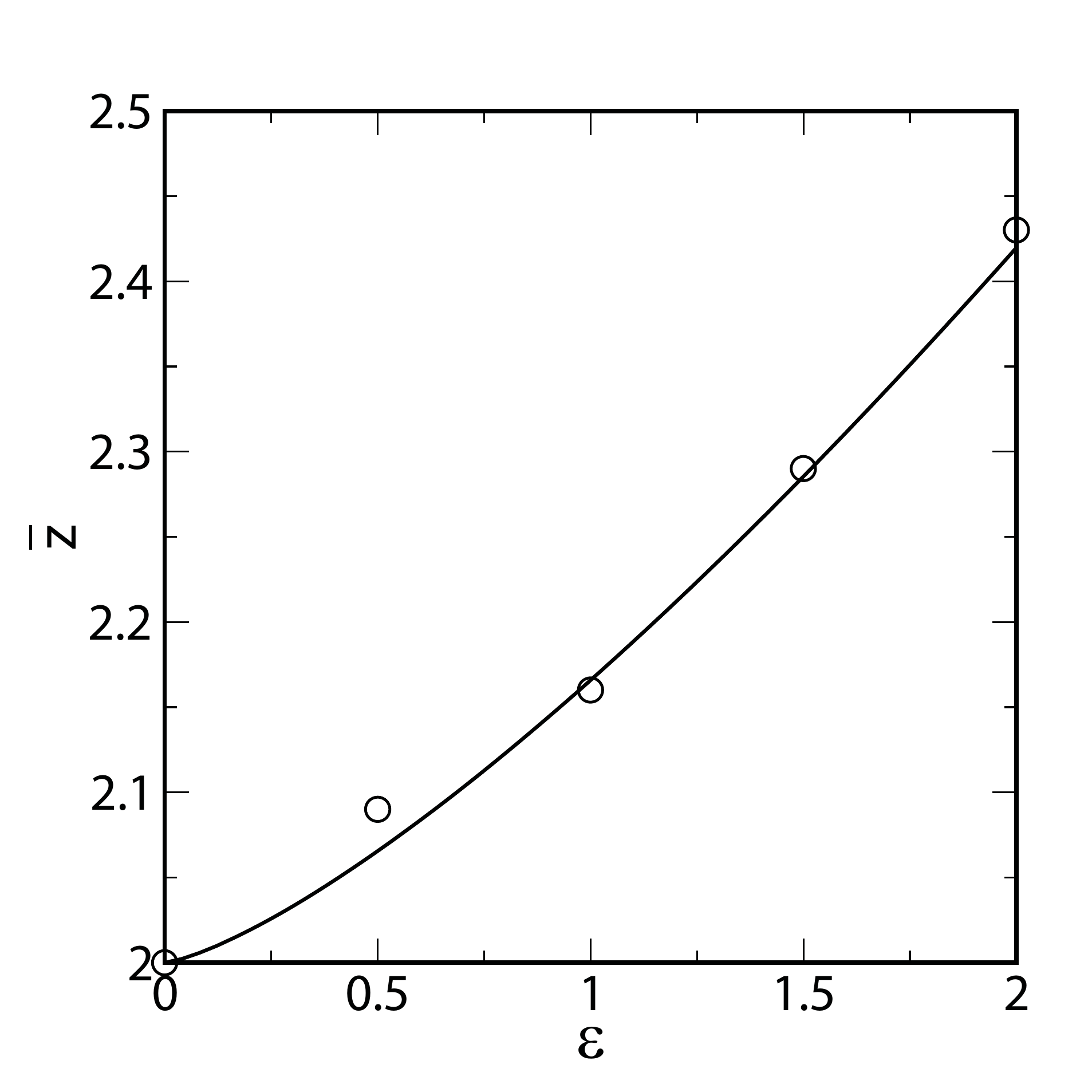}\\
\caption{Plot of the  disorder-dependent growth exponent $\overline{z}(\epsilon)$ vs. $\epsilon$. The solid line is the best power-law fit: $\overline{z}=2.0+0.166\epsilon^{1.34}$.}
\label{fig7}
\end{figure}   
 
In the early papers of Paul et al.~\cite{ppr} on the RBIM, it was argued that $\bar{z}_{\rm eff} $ scales linearly with the disorder amplitude $\epsilon$. This is a consequence of energy barriers which scale logarithmically with the length scale $R(t)$. In Fig.~\ref{fig7}, we plot $\overline{z}$ vs. $\epsilon$. We see that $\overline{z}$ increases somewhat faster than linearly \cite{henkel2006,henkel2008} -- a best fit to the data suggests $\overline{z}=2.0+0.166\epsilon^{1.34}$.

\subsection{RBXYM in $d=3$}

Next, we briefly present results for the $d=3$ RBXYM. As in the $d=2$ case, we first determine the transition temperature $T_c(\epsilon)$. Then, we study phase ordering kinetics after quenching the system below $T_c(\epsilon)$.     

\subsubsection{\emph{Estimation of $T_c$}}

\begin{figure}[htbp]
\includegraphics[width=\columnwidth]{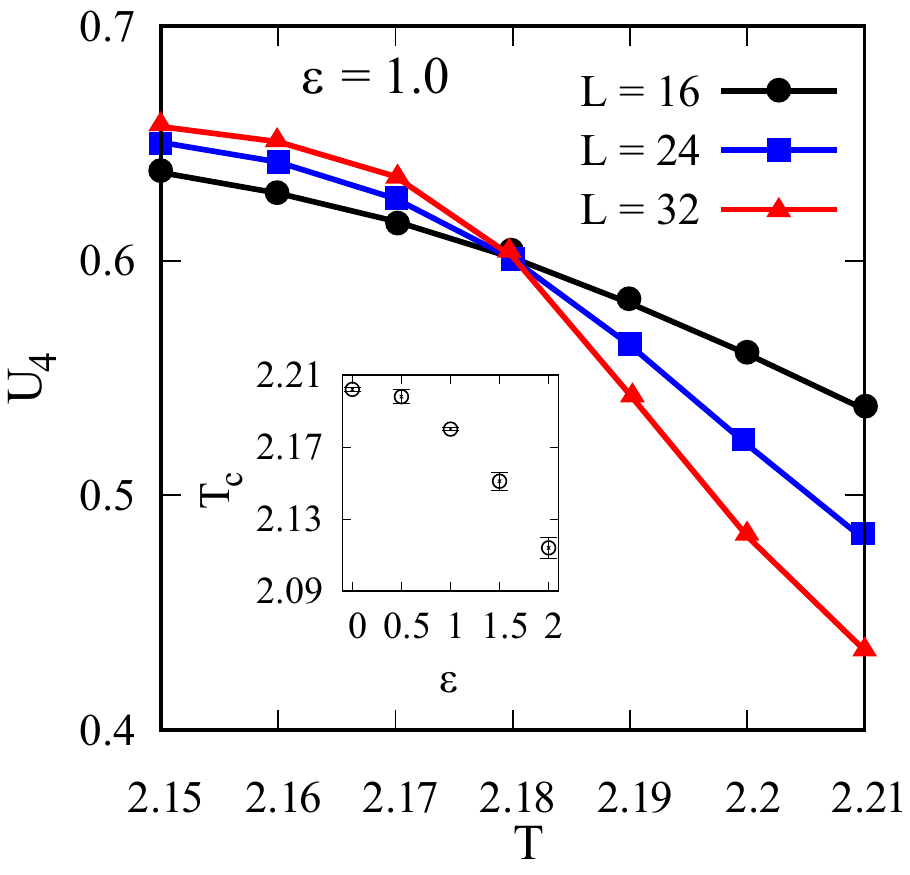}\\
\caption{Plot of $U_4(T,L)$ vs. $T$ for the $d=3$ RBXYM on cubic lattices of size $L = 16, 24,32$. We show data for $\epsilon = 1.0$. In the inset, we plot $T_c(\epsilon)$ vs. $\epsilon$. The numerical values of $T_c(\epsilon)$ are provided in Table~\ref{table2}.}
\label{fig8}
\end{figure}

To determine $T_c(\epsilon)$, we perform simulations on a simple cubic lattice of linear sizes $L$ = $16$, $24$ and $32$. After equilibration,  data for $m^2$ and $m^4$ are thermally averaged over $10^6$ MCS. We further average over 100 independent $\{J_{ij}\}$-configurations. Figure~\ref{fig8} shows the plot of $U_4$ vs. $T$ for $\epsilon=$ 1.0. As before, $T_c$ is determined from the intersection of Binder-cumulant curves for different $L$.  The $T_c$-values for various $\epsilon$ are given in  Table~\ref{table2}, and plotted in the inset of Fig.~\ref{fig8}. For the pure case ($\epsilon$ = 0), we obtain  $T_{c} \simeq 2.202$, which agrees with the values reported in the literature~\cite{hm,fmw}. With increasing $\epsilon$, $T_{c}$ decreases   to $T_{c}(\epsilon=2)\simeq 2.114$. For $d=3$,  $T_{c}$ does not change significantly with $\epsilon$, in contrast to the $d=2$ case.

\begin{table}[htbp]
\begin{center}
\begin{tabular}{|p{0.05\textwidth}|c|}
\hline
$\epsilon$ & $T_c$  \\
\hline
0.0 & 2.202 $\pm$ 0.001 \\
\hline
0.5 & 2.198 $\pm$ 0.004 \\
\hline
1.0 & 2.181 $\pm$ 0.004 \\
\hline
1.5 & 2.151 $\pm$ 0.005 \\
\hline
2.0 & 2.114 $\pm$ 0.006 \\
\hline 
\end{tabular}
\caption{Critical temperatures $T_c(\epsilon)$ for the $d=3$ RBXYM.}
\label{table2}
\end{center}
\end{table}

\subsubsection{\emph{Coarsening dynamics}}

Let us now discuss numerical results for domain growth in the $d=3$ RBXYM. The simulations are performed on  cubic lattices of size $128^3$. The system is quenched to $T = 1.0<T_c(\epsilon)$ at $t=0$ MCS, and evolved upto $t$ = $10^{6}$ MCS. The statistical data presented here is averaged over 10 independent realizations of disorder.

\begin{figure}[htbp]
\includegraphics[width=\columnwidth]{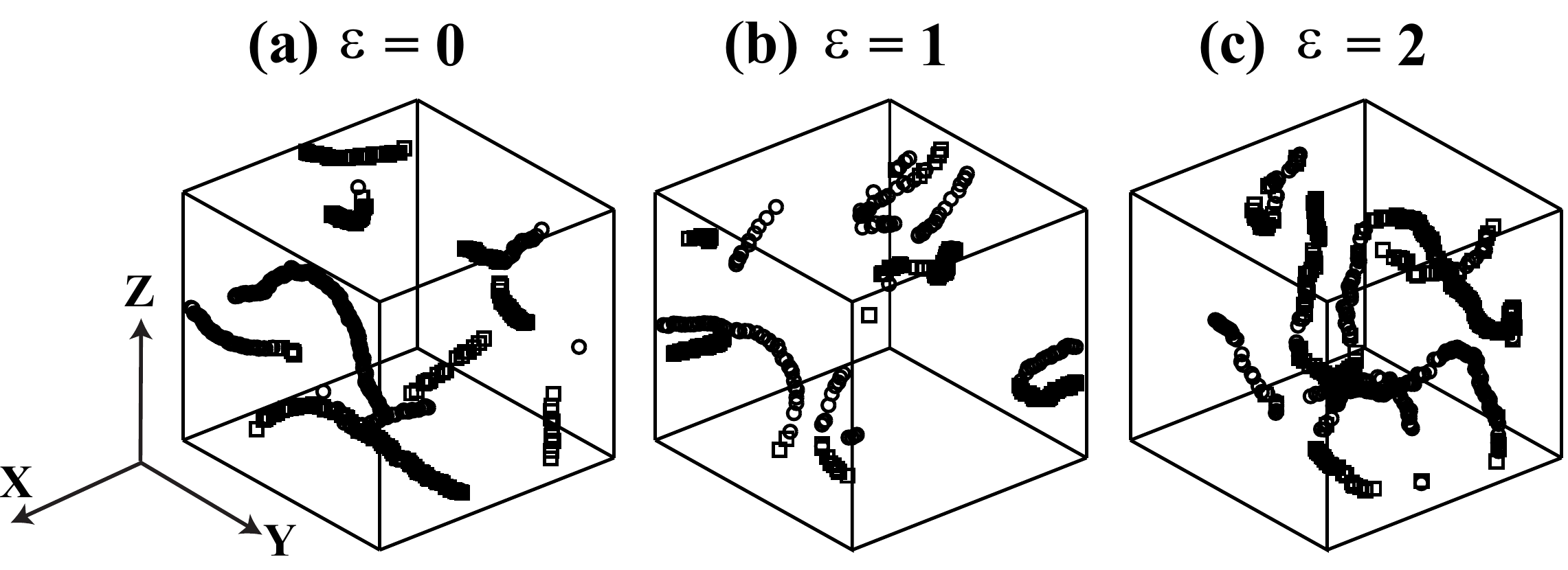}\\
\caption{Evolution snapshots of the $d$ = 3 RBXYM for a temperature quench to $T=0.5$. The lattice size is $128^3$. The defects (vortex and anti-vortex strings) are shown at $t$ = $10^{6}$ MCS, and for $\epsilon=0, 1, 2$.}
\label{fig9}
\end{figure}

Figure~\ref{fig9} shows the typical defect configurations for the evolution of the $d=3$ RBXYM system. The relevant defects in this case are vortex and anti-vortex strings. Here, we show the string configurations at $t=10^6$ MCS for $\epsilon=0,1,2$. The defect density reduces as the system evolves, due to the annihilation of vortices and anti-vortices. In Fig.~\ref{fig9}, we see that the defect density is higher for larger $\epsilon$, i.e., domain growth is slower for larger values of $\epsilon$.

\begin{figure}[htbp]
\includegraphics[width=\columnwidth]{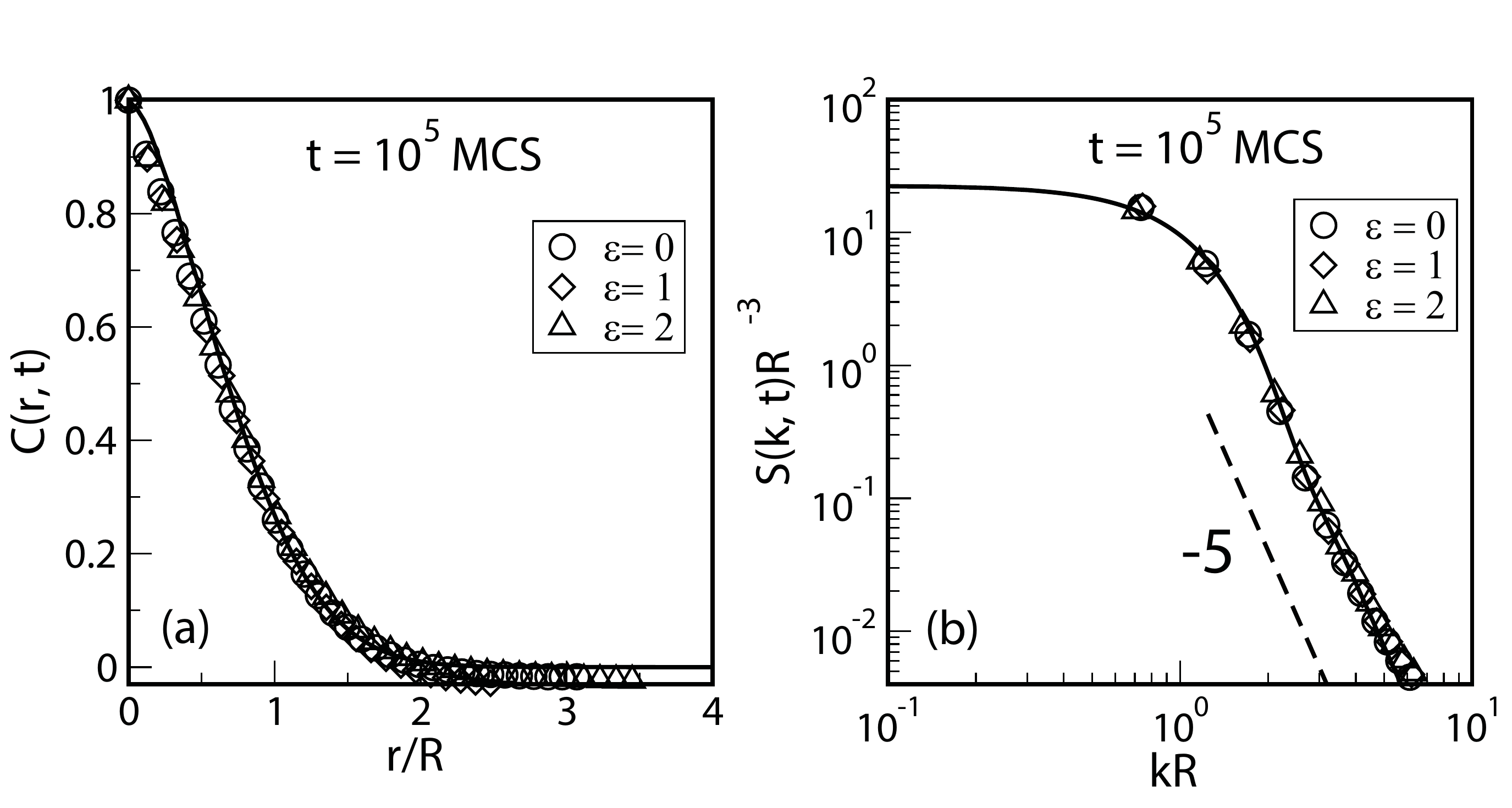}\\
\caption{(a) Scaled correlation functions, $C(r,t)$ vs. $r/R(t)$, for the $d=3$ RBXYM after a quench to $T=1.0$. We show data sets for $t$ = $10^{5}$ MCS, and $\epsilon=0,1,2$. (b) Scaled structure factors, $S(k,t)R(t)^{-3}$ vs. $kR$, corresponding to the data sets in (a). The solid curves in (a) and (b) denote the BPT function [Eq.~\eqref{bpt}] for $n = 2$ and its Fourier transform, respectively. The dashed line of slope $-5$ in (b) denotes the generalized Porod law: $S(k,t)$ $\sim$ $k^{-(d+n)}$ for $d = 3$ and $n = 2$.}
\label{fig10}
\end{figure}

Figure~\ref{fig10} shows the scaled correlation function [$C(r,t)$ vs. $r/R(t)$], and the scaled structure factor [$S(k,t)R(t)^{-3}$ vs. $kR$], for the evolution of the $d=3$ RBXYM. The data is shown for $t=10^{5}$ MCS, and for $\epsilon=0, 1, 2$. The solid lines in Fig.~\ref{fig10}(a) and (b) denote the BPT function for $n=2$,  and the corresponding Fourier transform, respectively.  In Fig.~\ref{fig10}(b), a dashed line of slope $-5$ denotes the generalized Porod law: $S(k, t)\sim k^{ -(d+n)}$ for $d = 3$ and $n = 2$. The data collapse in Fig.~\ref{fig10} is excellent, confirming the SU behavior of the scaling function in $d=3$, similar to the  $d=2$ case. However, this SU does not extend to the autocorrelation function, again as in the $d=2$ case. For the sake of brevity, we do not show this data here. 

The most important feature of the growth process is the domain growth law. Let us first understand the growth law for the pure XY model. In $d=3$, the defects  are of dimension 1, i.e., strings, as seen in Fig.~\ref{fig9}. Therefore, the defect pair energy $E_p(R)$ $\sim$ $R\ln(R/a)$, where we have included a factor of $R$ for the defect-core volume. Then, the driving force per unit defect-core volume is $F(R) \sim -\ln(R/a)/R$. The relation $v = \mu F$ yields
\begin{equation}
\frac{dR}{dt} \sim \frac{1}{R} ,
\end{equation}
so that $R(t) \sim t^{1/2}$. 

\begin{figure}[htbp]
\includegraphics[width=\columnwidth]{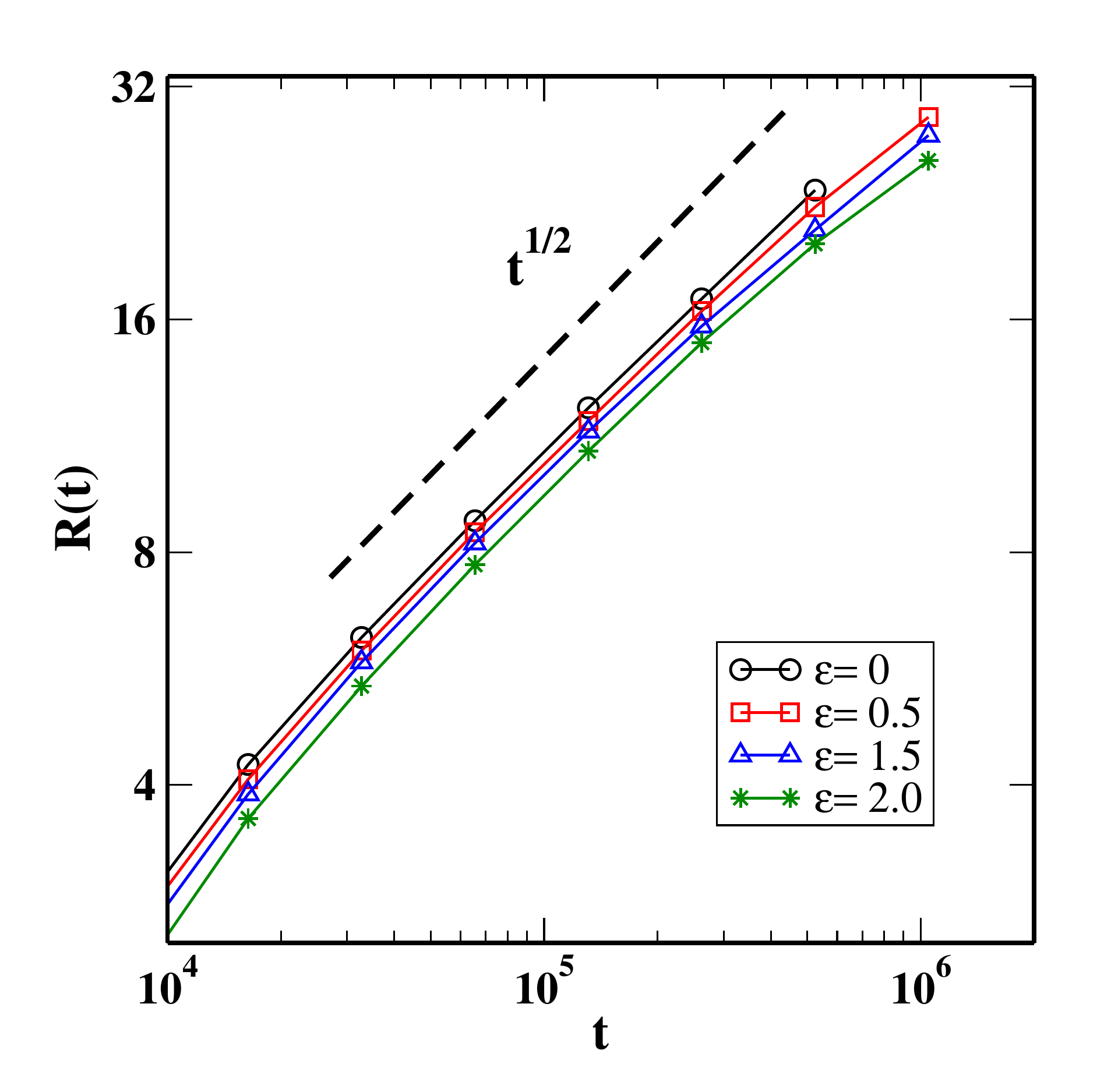}\\
\caption{Plot of $R(t)$ vs. $t$ (on a log-log scale) for the $d=3$ RBXYM. The dashed line denotes the $t^{1/2}$-growth for the pure case.}
\label{fig11}
\end{figure}

In Fig.~\ref{fig11}, we plot $R(t)$ vs. $t$ for $\epsilon= 0, 0.5, 1.5, 2.0$. The solid line denotes the growth law for the pure case: $R(t) \sim t^{1/2}$. We see that the growth law for the disordered cases follows the pure case for a while, and then becomes slower at late times. To understand the nature of the asymptotic growth law,  we define the effective exponent as
\begin{equation}
\frac{1}{ z_{\rm {eff}}} = \frac{d\,[\ln R(t)]}{d\, [\ln t]}.
\end{equation}

\begin{figure}[htbp]
\includegraphics[width=\columnwidth]{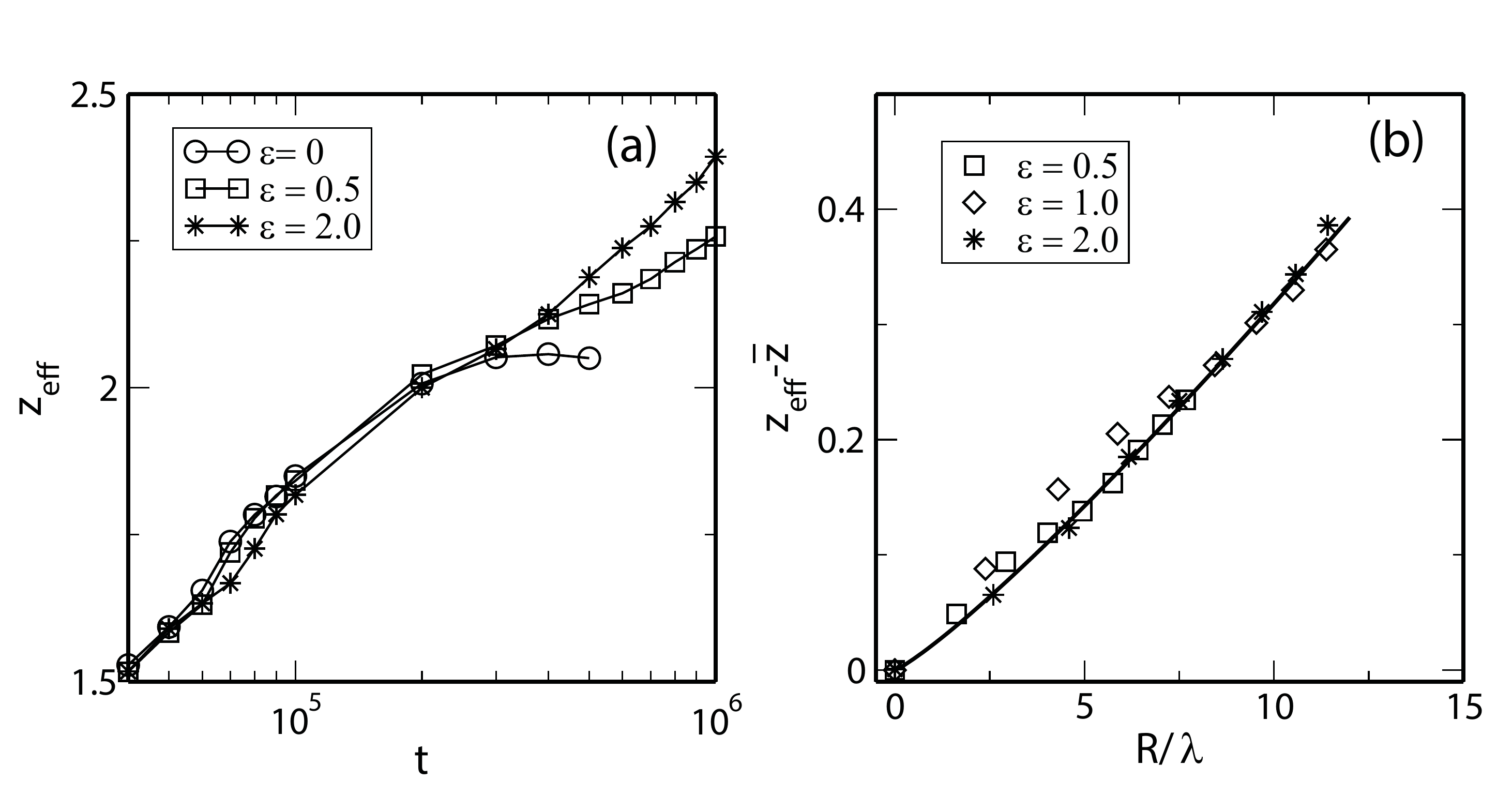}\\
\caption{(a) Plot of the effective exponent, $z_{\rm eff}$  vs. $t$. (b) Scaling collapse of $z_{\rm eff}-\overline{z}$ vs. $R/\lambda (\epsilon)$. The solid line is the best power-law fit:  $z_{\rm eff}-\overline{z} \simeq 0.022 (R/\lambda)^{1.16}$.}
\label{fig12}
\end{figure}

In Fig.~\ref{fig12}(a), we plot $z_{\rm eff}$ vs. $t$ for the data sets in Fig.~\ref{fig11}. At intermediate times, we have $z_{\rm eff}\simeq 2$, which is consistent with the pure case. At late times, the data for the disordered case shows an upward curvature, which has been understood by Corberi et al.~\cite{lmpz,clmpz2011,clmpz} as a signal of logarithmic growth. For each disordered data set, we identify $\overline{z}$ as the exponent value in the ``flat regime''.

In Fig.~\ref{fig12}(b), we plot $z_{\rm eff}-\overline{z}$ vs. $R/\lambda(\epsilon)$, where $\lambda(\epsilon)$ is a scaling variable. The late-stage data for different $\epsilon$-values shows a reasonable collapse. The scaling function is $z_{\rm eff}-\overline{z}\simeq b (R/\lambda)^\varphi$ with $b \simeq 0.022$ and $\phi \simeq 1.16$. As shown by Corberi et al.~\cite{clmpz}, this suggests the logarithmic growth law
\begin{equation}
\frac{R}{\lambda} \simeq \left[ \frac{\varphi}{b}\ln \left(\frac{t}{\lambda^{\overline{z}}}\right) \right]^{1/\varphi} , 
\end{equation}
with $1/\varphi \simeq 0.862$.

\section{Summary and Discussion}
\label{s4}

Let us conclude this paper with a summary and discussion of our results. We have undertaken a comprehensive Monte Carlo (MC) study of domain growth in the {\it random-bond XY model} (RBXYM) in $d=2,3$. Recall that the pure XY model exhibits the  {\it Berezinskii-Kosterlitz-Thouless} (BKT) transition in $d=2$, and shows a regular phase transition for $d=3$. To the best of our knowledge, this is the first study of the effects of quenched disorder on coarsening in systems with a vector order parameter. We find that the coarsening scenario observed in scalar systems by Corberi et al.~\cite{lmpz,clmpz2011} applies in the present case also.

We first summarize our $d=2$ study. We used the Binder-cumulant method to determine $T_{\rm BKT}$ as a function of the disorder amplitude $\epsilon$, where the random bonds $J_{ij}\in [1-\epsilon/2, 1+\epsilon/2]$. We only consider $\epsilon \leq 2$, so that there is no frustration in the system. We found that $T_{\rm BKT}$ decreased substantially as $\epsilon$ was increased. We then undertook a coarsening study by quenching an initially disordered system to $T<T_{\rm BKT}(\epsilon)$, where the equilibrium state is characterized by quasi-long-range-order (QLRO). Domain growth is characterized by the annealing of vortices and antivortices, which are point defects in $d=2$. The disorder sites trap these point defects and slow down coarsening. The evolution morphology is characterized by the {\it spatial correlation function} $C(r,t)$ and the {\it autocorrelation function} $A(t,t_w)$. The scaling form of $C(r,t)$ [or its Fourier transform, the {\it structure factor} $S(k,t)$] is not affected by the presence of disorder. However, this does not hold good for $A(t,t_w)$. On the time-scales of our simulation, the growth law exhibits a power-law behavior with an exponent which depends on the disorder amplitude. This is analogous to the intermediate-time behavior for coarsening in the RBIM~\cite{ppr,lmpz}. We do not see a logarithmic growth regime in the RBXYM, but expect it to arise on even later time-scales than those studied here ($t=10^6$ MCS).

Next, we summarize our $d=3$ study. In this case, the low-temperature [$T<T_c(\epsilon)$] state is characterized by LRO. Again, we use the Binder cumulant technique to determine $T_c(\epsilon)$. In $d=3$, the critical temperature does not show a strong dependence on $\epsilon$. Our results for domain growth in the $d=3$ RBXYM are analogous to those for $d=2$, with one major difference. In $d=3$, we do not see an extended intermediate regime of power-law  growth with an $\epsilon$-dependent exponent. Further, we see a clear signature of logarithmic growth in the asymptotic regime. 

Clearly, experimental systems always contain both quenched and mobile impurities. Therefore, it is important to have a good theoretical understanding of domain growth with quenched disorder. This now exists, as a result of our work and that of several other groups. However, experimental studies have not kept pace with these developments. We urge experimentalists to undertake careful experiments to confirm (or contradict) the theoretical scenario.

\section*{Acknowledgments}

SC thanks grant no. 09/080(0897)/2013-EMR-I, CSIR, India. RP thanks grant no. 03(1414)/17/EMR-II, CSIR, India.

\end{document}